# Quality assurance of e-learning processes

Zbigniew Mrozek
Faculty of Electrical and Computer Engineering
Politechnika Krakowska (Cracow University of Technology)
Krakow, Poland, Zbigniew.Mrozek@pk.edu.pl

*Abstract*—**A quality assurance system (QA) should ensure that student needs are met. It also respects accreditation requirements and student perceptions, supports training and development of teaching staff, controls costs and improves efficiency of e-learning system.**

*Keywords - eLearning; quality; ISO9000; ISO9004-2; PAS 1069; ELLEIEC project*

## I. Introduction

Over the past few years, with financial support of the European Union as well as the EU member states, numerous initiatives have been developed on QA (quality assurance) in education.

It is important to distinguish between QA and quality control. Contrary to quality control (testing the results), QA includes all activities needed to provide effective services for customers during the basic educational process and the full life cycle of the graduate

QA is currently perceived as something essential in Europe and internationally [3]. This may be regarded as an imperative for the European Commission, to take further its programmes and activities in the field of QA. Quality in e-learning has a twofold significance in Europe:

- First, decision-makers believe that e-learning will increase the quality of educational opportunities, ensuring that the shift to the information society is more successful. This 'quality through e-learning' is byond scope of this paper

- Second, it is a way to improve the quality of e-learning itself. This is known as 'quality for e-learning' and is described in this paper

## II. Sources of information about quality in e-learning:

Most information on QA may be found in Internet. There are also. books, standards, quality strategies, checklists and conference papers.

The most common official quality management approaches are ISO 9000 [5] family of standards, EFQM (European Quality Foundation Model, [2]), PAS [1] or quasistandards such as SCORM (Sharable Content Object Reference Model, [10]). Standards are also used by decision-makers in government, accreditation bodies, companies and universities.

## III. Service delivery characteristics in ISO 9004-2

The ISO9004-2 shows how to set up and manage a quality system that has a service orientation Any organization can benefit from following ISO's quality management recommendations. It describes needs for:

- Service accessibility and availability.
- Service safety, security, and reliability.
- Comfort and attractiveness of facilities.
- Service delay, duration, and delivery times.
- Service capacity and size of service facilities.
- Number of service providers and service tools.
- Service hygiene and service provider cleanliness.
- Competence and knowledge of service providers.
- Courtesy, attentiveness, and communication skills.
- Quantity and types of service supplies and materials.

According to ISO 9004, the quality service system is one that [6-9]:
- Ensures that customer needs are met (most important)
- Receives regular feedback from customers.
- Considers the social aspects of service delivery.
- Respects customer perceptions and opinions.
- Pays attention to the culture of the organization.
- Supports personnel training and development.
- Controls costs and improves efficiency

The above may be used to derive objectives of an educational institution
- satisfaction of customer (learner), consistent with the professional standards
- continuous improvement of educational service
- giving consideration to the requirements of industry, commerce and the public sector
- efficiency in providing the educational service

## IV. QUALITY OF E-LEARNING EDUCATION

There are three general aspects of quality [14]

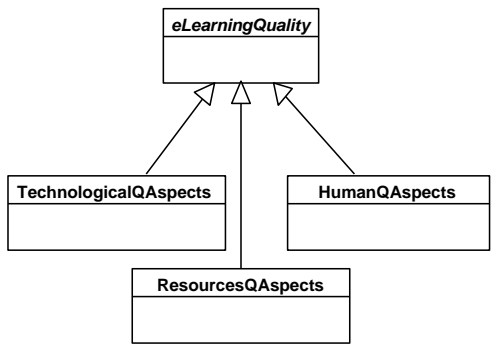

**Figure 1. General quality aspects in e-learning [14]**

### A. The technological qualty aspect

The technological aspect is related to computing environment, i.e.:

- design, implementation and development quality for e-learning systems, including the development of associated standards;
- adaptation and integration of computer technologies with existing e-learning systems and associated standards;;
- user expectations related to e-learning system, including support for personalization and customization, needs of learners related to unlimited access to materials, support for interoperability with other platforms, ease of use, work speed.
- ensuring proper security level in data protection and ensuring data recovery after failure,

### B. The e-resources qualty aspect

The e-resources aspect is related in didactic aspect to

- quality of e-learning materials,
- conformance to teaching model(s) [1, 2, 12, 16].

In non-didactic aspect to

- amount and quality of multimedia used,
- quality of e-resource development process.

### C. The human qualty aspect

In human aspect relates to two general groups of participants:

- direct participants, who are suppliers and designers of e-learning systems, teachers, methodology specialists, trainers and learners (students);
- indirect participants, who are authorities, accreditation and standardization bodies, law establishing and law regulating institutions, etc.

## V. SCORM, DE FACTO INDUSTRY STANDARD

SCORM is a powerful tool for anyone involved in creation of e-learning courses. Educational content can be used in many different systems and situations without modification. This plug-and-play functionality can be used within an organization or across organizations. Content can be sold and delivered to the user more quickly, more robustly, and at a lower price [12].

SCORM is widely adopted by some huge organizations. It is the de facto industry standard and appears in a vast majority of both training content and Learning Management Systems.

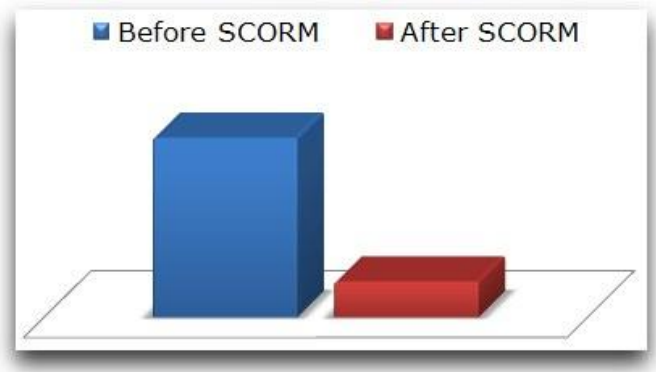

**Figure 2. The cost of content integration (http://scorm.com/scorm-explained/)**

## VI. QUALITY CONTROL AND DATA PROCESSING

The quality may be monitored with questionnaire for carrying out a user survey. Questionnaire has questions on actual e-learning module and on expectations of learners and mentors. The questionnaire asks about:

- general information about type of the user (learner or mentor), who is filling the questionnaire
- quality of learning contents available in actual module – as seen by actual user
- quality of graphical user interface
- features and quality of actual platform and actual e-resources
- effectiveness of e-learning
- features and contents that are desired but not available in actual module or in other courses

In order to perform statistical analysis of data gathered from the questionnaire, set of measures that characterized e-learning platforms should be constructed.

Classic methods fail for some input data. The possible new approach to data processing of questionnaires is to use data-mining technology, e.g. cluster analysis, the classification technique that can group objects according to their characteristics. Figure 3 shows results of using GCCA (Grade Correspondence Cluster Analysis) to investigate artificial data sets. [14, 15].

## VII. Quality Assurance in the VCE Project

The learner using VCE (Virtual Centre for Enterprise) should have a good educational experience with each module, such that he wants to return to take further modules. Friendly user interface helps to ease of access relevant content and activities.

Seligman [11] shows quality assurance measures, specifically useful for the improvement of distance learning,

- materials that are learner friendly, academically respectable, able to be used by the average student, interesting in content and layout, and relevant;
- learning materials and any peripheral media or equipment that are readily available;
- tutors and students that become familiar with distance learning methodology and practice;
- the whole system that is managed effectively;
- monitoring, evaluation, and feedback that are viewed as important

During the design and development phase careful consideration is given to the student journey through each module and what is required by the system (prerequisite) at each stage of this journey. This detailed analysis generated a list of requirements ranging from learner briefings, training documents, mentor specifications, mentor briefings, module specification templates, eg [17, 18, 19]:

*1)* clear definition of learner needs with appropriate quality measures

*2)* preventive action and controls to avoid customer dissatisfaction

*3)* optimising quality-related costs for the required performance and grade of service

*4)* creation of collective commitment to quality within the educational organisation

*5)* continuous (and never ending) review of service requirements and achievements to identify opportunities for service quality improvement

To build a deeper understanding of pedagogy and assessment issues, a literature resource has been built into the VCE so that module designers and mentors can easily find and gain access to relevant published material.

Following guidelines should shape the quality of e-learning [18]:
- learners must play a key part in determining the quality of e-learning services;
- quality must play a central role in education and training policy
- quality must not be the preserve of large organisations;
- support structures must be established to provide service-oriented quality assistance;
- open quality standards must be further developed and widely implemented;
- appropriate business models must be developed for services in the field of quality

Objectives to be discussed:
- to investigate the use of quality approaches: how are they used?
- to identify possible factors for success, on which the development of quality may depend.

The quality development is a competence that must be possessed by those involved in the learning process – in e-learning, for example, by tutors, media designers, authors and of course learners – if successful quality development is to be made possible. This competence can be broken down into four dimensions [17, 18, 19]:

(a) knowledge of what opportunities are available for quality development;

(b) ability to act and experience of using existing quality strategies;

(c) ability to adapt and further develop, or to design original quality strategies;

(d) critical judgement and analytical ability to enhance quality in one's own field of operation.

In order to achieve a successful survey, a large a number of people involved in e-learning should be reached and a broad spectrum of e-learning experts, e-learning decision-makers and e-learning users shouls be included.

## VIII. Conclusions

- The introduction of QA methodology should be considered by all educational institutions.
- Contrary to quality control (testing the results), QA includes all activities needed to provide effective services for customers during the basic educational process and the full life cycle of the graduate
- A graduate is a product of an educational institution.
- Modern tools and technology should be used in designing and validation of e-learning systems – see Fig. 2, Fig. 3 and Fig. 4.

QA of e-learning should involve all participants. Quality is influenced both by qualifications of a team designing a course and teachers who realize it. One cannot also forget the degree of involvement of teachers, learners and facilities available in the learning processes.

### Acknowledgment

The author wish to thank the European Commission for the grant to ELLEIEC in the Life Long Learning Programme and all the partners of the project for their contribution.

Project number is 142814-LLP-1-2008-FR-ERASMUS-ENW, and instrument type is: ERASMUS NETWORK.

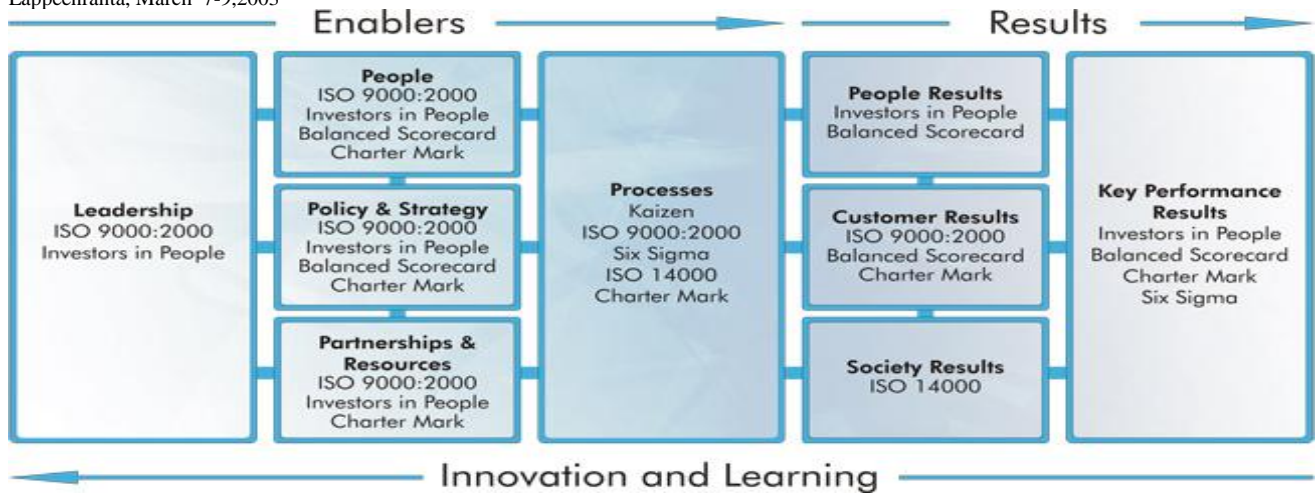

**Figure 4 The Excellence Model of EFQM, http://www.qualityscotland.co.uk/assets/page%20graphics/efqm-model.jpg**

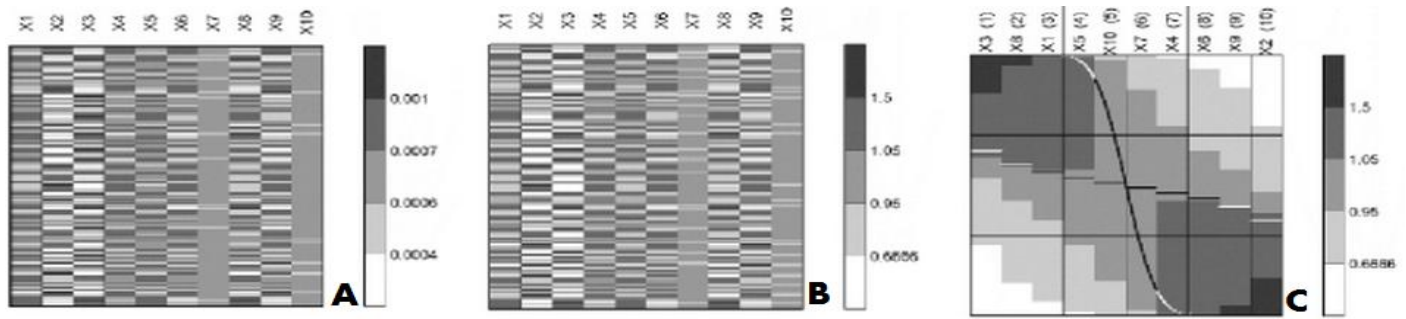

**Figure 5. Using datamining tools: Grade Correspondence Cluster Analysis (GCCA) to investigate artificial data sets. (A) raw data maps, (B) overrepresentation of nonsorted rows and columns of data matrix, (C) overrepresentation map for rows and columns sorted and clustered according to GCCA [14, 15].**